\documentclass[11pt]{article}
\usepackage[margin=1in]{geometry}
\usepackage{graphicx}

\usepackage{physics}
\usepackage{latexsym}
\usepackage{amsmath}

\def\beq{\begin{equation}}
\def\eeq#1{\label{#1}\end{equation}}
\def\eeqn{\end{equation}}

\def\beqa{\begin{eqnarray}}
\def\eeqa#1{\label{#1}\end{eqnarray}}
\def\eeqan{\end{eqnarray}}

\let\bar=\overbar

\def\ket#1{\left| {#1} \right\rangle}

\def\Dslash{\not{\hbox{\kern-4pt $D$}}}
\def\dslash{\not{\hbox{\kern-2pt $\del$}}}

\def\msb{{\bar{\ssstyle M \kern -1pt S}}}

\def\Title#1{\begin{center} {\Large {\bf #1} } \end{center}}
\def\Author#1{\begin{center} {\normalsize {\sc #1} } \end{center}}
\def\Institution#1{\begin{center} {\normalsize {\it #1} } \end{center}}
\def\Abstract#1{\noindent {\normalsize {\bf Abstract:} {\normalfont #1}}}
\def\Conference{\vspace{4mm}\begin{raggedright} {\normalsize {\it Talk presented at the 2019 Meeting of the Division of Particles and Fields of the American Physical Society (DPF2019), July 29--August 2, 2019, Northeastern University, Boston, C1907293.} } \end{raggedright}\vspace{4mm}}

\begin{document}

%
%

\Title{Novel matter effects on neutrino oscillations observables}

\Author{Adam Zettel and Mihai Horoi}

\Institution{Department of Physics, Central Michigan University, Mount Pleasant, Michigan 48859, USA}

\Abstract{In a recent article \cite{mh-msw18} we noticed that the electron density in condensed matter exhibits large spikes close to the atomic nuclei. We showed that these spikes in the electron densities, 3-4 orders of magnitude larger than those inside the Sun's core, have no effect on the neutrino emission and absorption probabilities or on the neutrinoless double beta decay probability. However, it was not clear if the effect of these spikes is equivalent to that of an average constant electron density in matter. We investigated these effects by a direct integration of the coupled Dirac equations describing the propagation of flavor neutrinos into, through, and out of the matter. We found little evidence that these spikes affect the standard oscillations probabilities, but found a new fast and efficient algorithm of calculating these probabilities for neutrinos propagating through varying electron densities.}

\Conference

%
%

\section{Introduction} \label{intro}

The results of the solar and atmospheric neutrino oscillation experiments were recognized by a recent Nobel prize. The Mikheyev-Smirnov-Wolfenstein (MSW) effect is an essential component needed for the interpretation of these neutrino oscillation experiments \cite{16smirnov} (for a historical account of neutrino oscillations in matter see \cite{19smirnov}).  Therefore, the mixing of the neutrino mass eigenstates in vacuum and in dense matter seem to be well established in describing the propagation of the neutrinos from source to detecting devices. These effects were mostly considered in electron plasma \cite{haxton86,parke86}, such as the Sun or supernovae, and its use is extended to condensed matter, such as the Earth crust and inner layers. However, to our best knowledge, the variation of the electron density inside condensed matter was not yet considered. A simple estimate of the electron density and neutrino potential inside a medium-Z nucleus, such as $^{136}$Xe, shows that it is about four orders of magnitude larger than that existing in the Sun's core \cite{mh-msw18}. One could then ask if these high electron densities can produce additional mixing of the mass eigenstates that needs to be considered in the interpretation of neutrino production and detection phenomenology.

The effect of the matter-induced neutrino potential on the neutrino mixing in matter is traditionally analyzed using the local in-medium modified mass eigenstates and mixings \cite{giunti2007,88mannheim1935,92giunti1557}. This approach relies on the separation of the neutrino wavelength scale from the much larger neutrino oscillation and matter density variation scales. In reality, there are no local mass eigenstates in matter, but the analysis of the evolution of the vacuum mass eigenstates in finite matter medium is complicated by the various scales involved. However, the results based on local, in-medium, mass eigenstates seem to be valid. One of the issues related to the introduction of the fictitious in-matter mass eigenstates is that one assumes that the neutrinos produced via weak interactions in dense matter (e.g. in the Sun's core) are emitted as matter mass eigenstates. The transition to vacuum is usually described by the long-scale evolution of the amplitudes (see e.g. Ref. \cite{03garcia345}), which could be adiabatic or not. We are investigating if this approach can be extended to the analysis of the effects of non-adiabatic transitions of the neutrinos through condensed matter
where the electron densities near the atomic nuclei are few orders of magnitude larger than this in the Sun's core.

\section{Neutrino oscillations in condensed matter} \label{spikes}

\begin{figure}[htb]
\centering
\includegraphics[height=3.5in]{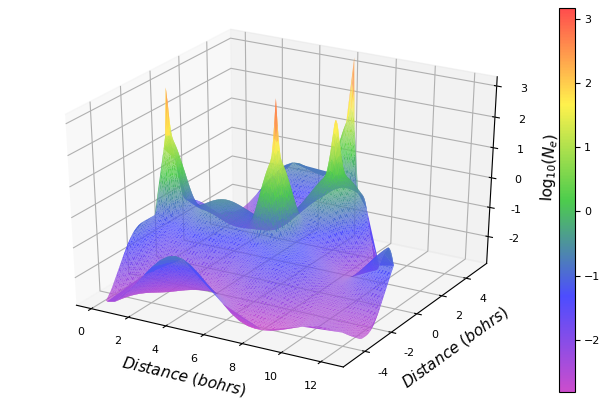}
\caption{Electron density inside a quartz ($SiO_2$) cell obtained with DFT calculations using Quantum Expresso code \cite{qe2009}. Shown is the electron density (in atomic units) in a plane through the cell that cuts very close to three silicone nuclei (higher peaks), and two oxygen nuclei.}
\label{fig:sio2}
\end{figure}

It is now widely accepted that the flavor neutrinos participating in the weak interaction are coherent superpositions of vacuum mass eigenstates. For the neutrino fields, the mixing reads:
\begin{equation} 
\nu_{\alpha L}(x)=\sum_{a=1} U_{\alpha a} \nu_{aL}(x)\ ,
\label{mee}
\end{equation}
where index $\alpha$ indicates a flavor state (electron, muon, tau, $\ldots$), and $a$ designates mass eigenstates (1, 2, 3, $\ldots$), and $U_{\alpha a}$ are elements of the vacuum neutrino mixing (PMNS) matrix. Here the dots indicate sterile flavors, or high mass eigenstates. If one discards the existence of the low mass sterile neutrinos,  the coupling to the higher mass eigenstates is then very small, and the sum over $a$ in  Eq. (\ref{mee}) is reduced to 3. This mixing leads to violations of the flavor number, and it is reflected in the outcome of  the neutrino oscillation experiments. These experiments are mostly  analyzed in terms of neutrino states
\begin{equation} \label{kvm}
\ket{ \nu_{\alpha L}}  = \sum_{a=1} U_{fa}^* \ket{\nu_{aL}}\ ,
\end{equation}
which are dominated by the larger components of the fields. Neutrino states are used to analyze the matter effects, also known as Mikheyev-Smirnov-Wolfenstein effects \cite{16smirnov}. 
Neutrino mixing is affected in matter by the neutrino optical potential.
The general relation between the neutrino optical potential (in $eV$) and the electron density $N_e$ (in $cm^{-3}$) is 
\begin{equation} \label{vmat}
V_e(eV)  = \pm \sqrt{2}G_F N_e \approx \pm 7.56\times 10^{-14}  m_p N_e =\pm 1.26\times 10^{-37} N_e \ ,
\end{equation}
where the (minus)plus sign corresponds to (anti)neutrinos, $G_F$ is Fermis's constant, and $m_p$ is the proton mass ($1.67 \times 10^{-24}  g$). Above we used Eq. (2.8) of \cite{92giunti1557}, where the equivalent matter density times the electron fraction $Y_e$ was replaced with $m_p N_e$.
In atoms, just considering the electron density of two electrons in the lowest $1s$ state of a Hydrogen-like atom (the higher s-states contribute very little, $\propto 1/n^3$, $n$ being the principal quantum number), one gets
\begin{equation} \label{rhot}
 N_e(r) =  10^{30} \frac{2}{\pi}\left(\frac{Z}{53}\right)^3 e^{-2 r Z/53}\ (cm^{-3}),
\end{equation}

\noindent
where $Z$ is the atomic number, and $r$ is in $pm$ ($10^{-12}\ m$). Electron DFT calculations \cite{dftprl1,dftprl2} (see e.g. Fig. 1 of Ref. \cite{mh-msw18}), show that this approximation is very good at and near the nuclei, where the main transition takes place. Fig. \ref{fig:sio2} shows the result of a DFT calculation for the electron density in the cell of quartz crystal, one of the most common in the Earth's crust. The results show that 85\% of the electron density in the cell resides in the spikes (defined as larger than the average density of 0.1 atomic units), while only 15\% is located in the volume that has lower than average density. In addition, the values of the electron density near the peaks are very well described by Eq. (\ref{rhot}).
These high electron densities near the nuclei are much larger than those in the Sun's core for all atoms with atomic number greater than 5. As an example, for atoms with $Z \approx 53$ the electron density at the nucleus  is four orders of magnitude larger than that in the Sun's core.

Therefore, it should be interesting to investigate the effects of these large electron densities on the neutrino mixing in atomic weak interactions. To solve this problem one needs to consider the evolution of mixing for three (or more) neutrino mass eigenstates, which can be described by the coupled Dirac equations
\small
\begin{equation} \label{aevolved}
i\frac{d}{dt}
\begin{pmatrix} 
\psi_{1} \\
\psi_{2} \\
\psi_{3}
\end{pmatrix} = 
\left[
\begin{pmatrix} 
p_x \alpha_x + m_1 \beta & 0 & 0 \\
0 & p_x \alpha_x + m_2 \beta & 0  \\
0 & 0 & p_x \alpha_x + m_3 \beta  
\end{pmatrix}
+ U^{\dagger}
\begin{pmatrix} 
V_e(x)+V_N & 0 & 0 \\
0 & V_N & 0  \\
0 & 0 & V_N   
\end{pmatrix}
U
\right]
\begin{pmatrix} 
\psi_{1} \\
\psi_{2} \\
\psi_{2}
\end{pmatrix}\ ,
\end{equation}
\normalsize
where $\psi_i$ are Dirac spinors for the (perturbed) vacuum mass eigenstates, $m_i$ are the corresponding neutrino masses, $p_x$ is the momentum in the direction of the beam, and $\alpha_x$ and $\beta$ are Dirac matrices \cite{peskin1995}. 
Here $V_N$ is the neutral current potential generated mostly by neutrons, i.e.
\begin{equation} \label{VN} 
V_N(eV) \approx - G_F N_n/\sqrt{2} \approx -6.3 \times 10^{-38} N_n \ ,
\end{equation} 
where $N_n$ is the local neutron density in $cm^{-3}$. The neutral current potential is the same for all three active neutrinos, and therefore can be neglected in the analysis of neutrino oscillations (as is the main momentum term in the underlying Dirac equations). However, if the sterile neutrinos are present, the neutral current potential needs to be considered \cite{parke19}. 

In Eq. (\ref{aevolved}) the Dirac spinors, $\psi_a = \nu_a \phi_a$ can be viewed as components of some flavor neutrino normalized superposition of mass eigenstates spinors,
\begin{equation}
\psi_{\alpha} = \sum_{a=1,2,3} \psi_a=\sum_{a=1,2,3} \nu_a \phi_a \ ,
\end{equation}
where $\alpha$ indicates the known active neutrino flavors, electron, muon, and tau. Here we consider the traditional approach \cite{03garcia345} of separating the neutrino wavelength scale from the neutrino oscillation and matter density variation scales, by considering a Schroedinger-like equation for the amplitudes, assuming that the $\phi_i$ spinors are free spinors,

\begin{figure}[htb]
\centering
\includegraphics[height=3.5in]{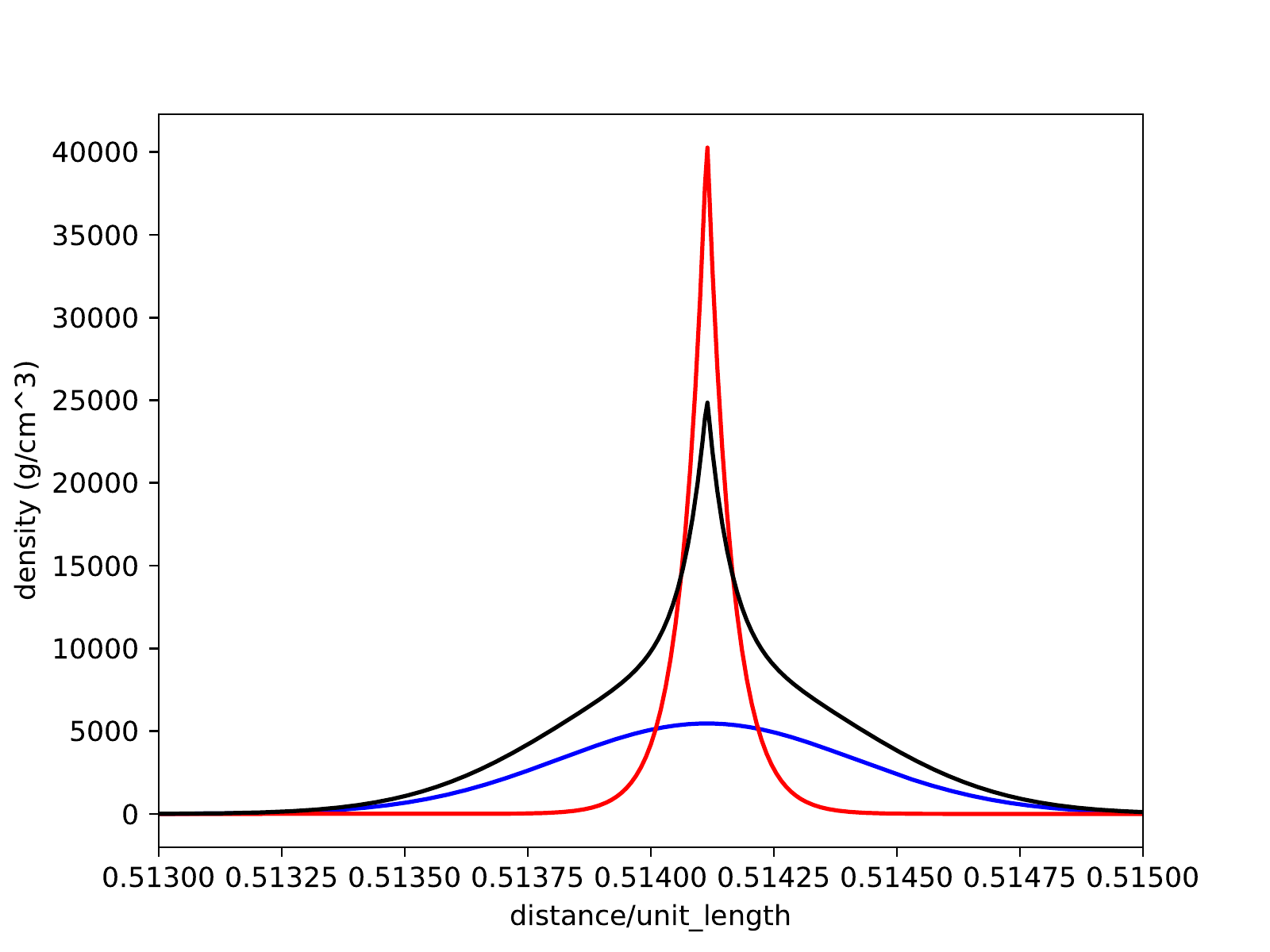}
\caption{Profile of the electron density spikes used in the calculations multiplied by the proton mass, for comparison to equivalent matter density, $Y_e \rho_{equiv}(s) = N_e(s) m_p$.} 
\label{fig:den-spikes}
\end{figure}

The vector of 3 flavor amplitudes is denoted as $\nu_f = \left( \nu_e,\ \nu_{\mu},\ \nu_{\tau} \right)^T$, and
then the Schroedinger-like evolution equation for the flavor amplitudes in matter reads
\begin{equation} \label{wolf}
i\frac{\partial \nu_f}{\partial t} = \left( H_0 + V \right) \nu_f \ ,
\end{equation}
where $H_0 = U diag \left( m_1^2/(2P), m_3^2/(2P), m_3^2/(2P)\right) U^{\dagger}$, $V = diag \left( V_e+V_N, V_N, V_N \right)$, and $m_a$ are the masses of the vacuum mass eigenstates. 
The general requirement for the validity of 
the above evolution equation is that the neutrino wavelength be smaller than the length over which there is a significant change of the optical potential created by a varying electron density  \cite{03garcia345,92giunti1557}, 
\begin{equation} \label{vcond}
\lambda \ll \mid V(x)/ \left( dV/dx \right) \mid \ .
\end{equation}
In the case of the potential created by the atomic electron density, Eqs. (\ref{vmat}) and (\ref{rhot}), this condition reads
\begin{equation}
2\pi\frac{\hbar c}{Pc} \ll \frac{53000}{2Z}\ \ (in\ fm),
\end{equation}
which is satisfied for neutrino energies larger than 2-5 MeV, and for a wide range of atomic numbers. 

In constant electron density Eq. (\ref{wolf}) is usually solved by diagonalizing the in-matter Hamiltonian, $H=H_0+V_e$, assuming that the solution describes in-matter mass eigenstates that have in-matter mixing matrix and masses, and using these effective masses and mixings in the standard vacuum oscillation formulae \cite{03garcia345}. We will call this the eigenvalues method. In this approach,
Eq. (\ref{rhot}) suggests that the electron density inside the atomic nucleus is much larger than that in the Sun's core and, therefore, the (anti)neutrinos are emitted in the (lower)higher mass eigenstates. Solutions to this problem for one single atom were discussed in Ref. \cite{mh-msw18}. Here we are interested to see if there are any effects of the electron density "spikes" around the atomic nuclei in bulk condensed matter.
For that we integrate Eq. (\ref{wolf}), which we rewrite in dimensionless form,
\begin{equation} \label{propfl}
i\frac{d}{ds}
\begin{pmatrix} 
\nu_{e} \\
\nu_{\mu} \\
\nu_{\tau}
\end{pmatrix} = 
\left[
U
\begin{pmatrix} 
0 & 0 & 0 \\
0 & \alpha & 0  \\
0 & 0 & \gamma  
\end{pmatrix}
U^{\dagger}+
\begin{pmatrix} 
A(s) & 0 & 0 \\
0 & 0 & 0  \\
0 & 0 & 0   
\end{pmatrix}
\right]
\begin{pmatrix} 
\nu_{e} \\
\nu_{\mu} \\
\nu_{\tau}
\end{pmatrix}\ \equiv H(s)
\begin{pmatrix} 
\nu_{e} \\
\nu_{\mu} \\
\nu_{\tau}
\end{pmatrix}\ .
\end{equation}
Here $s$ is the normalized propagation length, $s=x/s_u$, $s_u$ is the unit length defined as $s_u = (2E \hbar c)/|\delta m_{31}^2|$, $\alpha =  \delta m_{21}^2/|\delta m_{31}^2|$, $\gamma=\delta m_{31}^2/|\delta m_{31}^2|$, and $A(s)=2E V_e(x)/|\delta m_{31}^2|$. 
In the calculation we use a number $N$ of density spikes, entering $A(s)$ via $V_e(x)$. Shown in Fig. \ref{fig:den-spikes} are different density profiles: Gaussians in blue, exponential in red, and a combination of the two of them in black. Each density profile represents the electron density near an atomic nucleus multiplied by twice the proton mass. Given that the Gaussians are normalized to unity, an additional normalization factor $\rho_N$ was used to recover the flat average density of matter $\rho$,
\begin{equation} \label{norm_r}
\rho_N = \rho\ s/N \ ,
\end{equation}
where, for an easier comparison with solar density, we used the local equivalence between the average electron density $\left< N_e \right>$ and average mass density $\rho$,
\begin{equation} \label{rho_ave}
\rho = m_p \left< N_e \right>/Y_e \approx 2 m_p \left< N_e \right> \ .
\end{equation}

\begin{figure}[htb]
\centering
\includegraphics[height=3.5in]{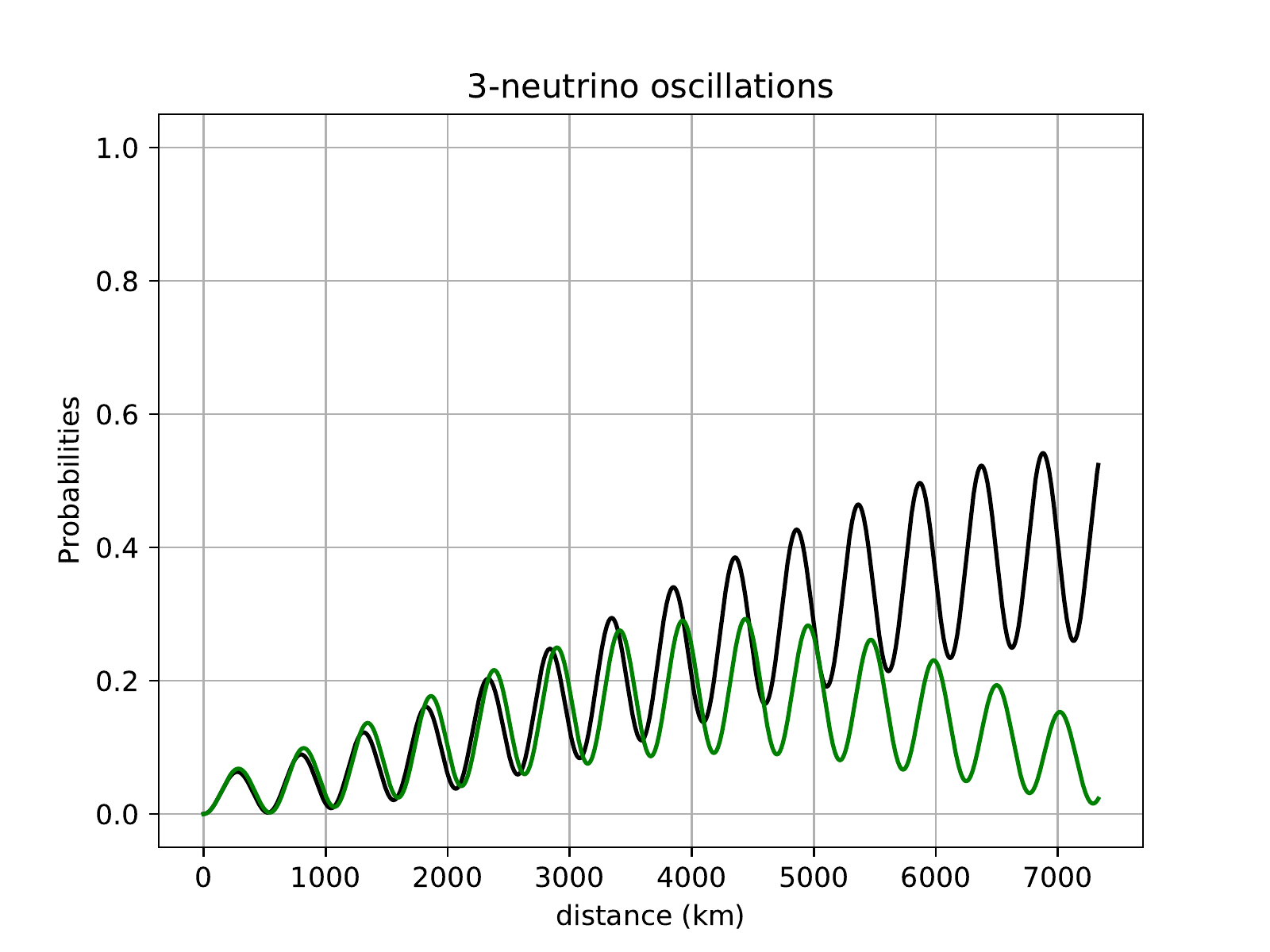}
\caption{Electron neutrino appearance probability, $P_{\nu_{\mu} \nu_e}$, for a baseline of 7330 km. The energy of muon neutrino beam is 0.5 GeV. The green line is given by a constant matter density approach ($\rho=3.8\ g/cm^3$), and the black line was obtained integrating Eqs. (\ref
{wolf}) using a spiked electron density (see text for details).}
\label{fig:pme}
\end{figure}

To integrate Eq. (\ref{wolf}) we used the ZVODE routine from the SciPy ODE package, which implements a complex version of the VODE algorithm \cite{vode89}. Given that the electron density spikes are extremely narrow, we tried different widths for the density profiles shown in Fig. \ref{fig:den-spikes}. In an attempt to get good accuracy of the solution we divided the width of each density spike by about 100 integration steps, and required for each step an absolute tolerance of $10^{-10}$ from ZVODE routine. The results in Fig. \ref{fig:pme} show a significant difference between the solution of the integration method using the spiked density profile (in black) and the``exact" eigenvalues solution corresponding to the equivalent flat electron density (in dark green). However, when increasing the accuracy in the ZVODE routine to $10^{-12}$ the difference between the two curves in Fig. \ref{fig:pme} disappeared. This situation emphasized once again the danger of relying solely on numerical analysis. In addition, the needed accuracy of $10^{-12}$ being close to the numerical round-off error for double precision, further emphasizes the difficulty of the numerical problem. Therefore, we tried using a more direct analysis to understand this result, which will be presented in the next section.

\section{Fast and efficient algorithm to calculate the neutrino oscillation probabilities through varying electron densities} \label{iters}

Previous work on neutrino oscillation probabilities in matter includes perturbative expansions \cite{parke16-1,parke16-2,parke18,parke19}. The mostly used approach is to diagonalize the Hamiltonian in matter, and use perturbation theory to identify main  contributing terms. In the process, one uses the S-matrix approach for the propagation of the amplitudes. For example, for the Eq. (\ref{propfl}) the corresponding S-matrix is given by,
\begin{equation}
S(s)=T e^{-i \int_0^s H(s')ds'} \ ,
\end{equation}
where the $T$ operator in front of the exponential indicates an ordered position-dependence of the integrals when the matrix exponential is expanded (similar to the time ordered product).
The S-matrix can be used to calculate the probability of measuring neutrino flavor $\alpha$ at distance $s$ assuming that flavor $\beta$ was emitted at $s=0$ \cite{parke16-1},
\begin{equation} \label{prob_s}
P_{\beta \rightarrow \alpha} = \left| S_{\alpha \beta} (s) \right|^2 \ .
\end{equation}
In case of constant electron density, $A(s)$ in Eq. (\ref{propfl}) does not depend of the integration variable $s$, and one can use the diagonalization methods or/and perturbation expansions \cite{parke16-1,parke16-2,parke18,parke19}. Alternatively, one can directly integrate Eq. (\ref{propfl}).

\begin{figure}[htb]
\centering
\includegraphics[height=3.5in]{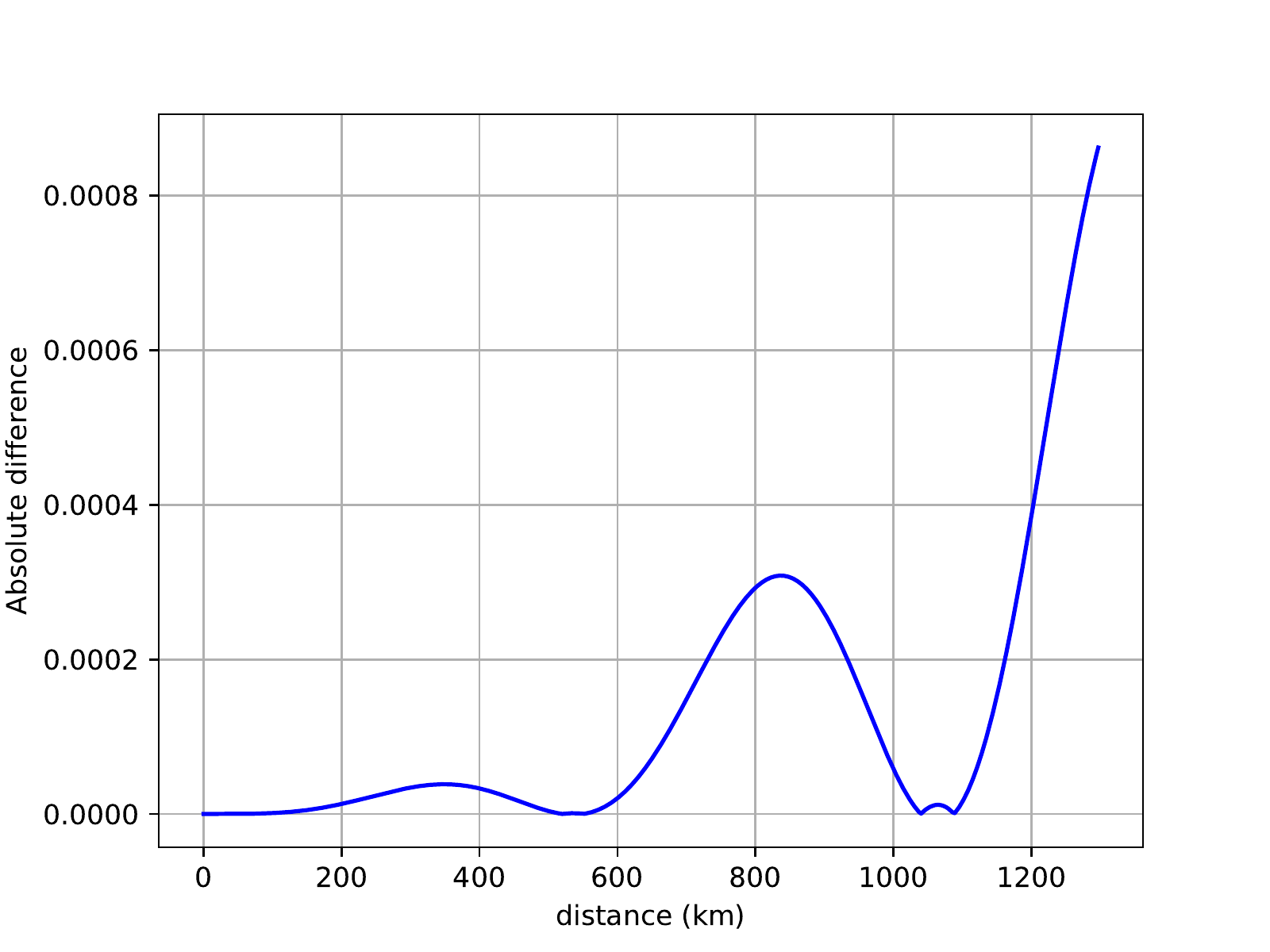}
\caption{Absolute difference between the probability of electron neutrino appearance calculated with the iteration method described in the text and the ``exact" eigenvalues method. Used was a matter constant density of $2.8\ g/cm^3$, and a muon neutrino beam of 0.5 GeV.}
\label{fig:errd}
\end{figure}

Here we propose using the matrix solution to Eq. (\ref{propfl}) in a different way: we divide the $s$ interval in $N$ small pieces $\Delta s_i$ (for example equally spaced $\Delta s_i = s/N$), for which we consider the $H(s_i)$ Hamiltonian constant. With the notation
\begin{equation}
H(s) \equiv U D_1 U^{\dagger} + D_2(s) \ ,
\end{equation}
the solution to Eq. (\ref{propfl}) can be written as
\begin{equation} \label{alg_p}
S (s) = \prod_{i=1}^N S (\Delta s_i) \ .
\end{equation}
Given that the $\Delta s_i$ are small, one can show that the matrices $S(\Delta s_i)$ can be approximated by
\begin{equation} \label{alg_s}
S(\Delta s_i) = e^{-i \Delta s_i D_2(s_i)} U e^{-i \Delta s_i D_1} U^{\dagger} \ .
\end{equation}
Moreover, given that matrices $D_1$ and $D_2$ are diagonal then,
\begin{equation} \label{alg_ua}
  e^{-i \Delta s_i D_2(s_i)} = 
\begin{pmatrix} 
e^{-i \Delta s_i A(s_i)} & 0 & 0 \\
0 & 1 & 0  \\
0 & 0 & 1   
\end{pmatrix} 
\equiv U_A(s_i) \ ,
\end{equation}
and
\begin{equation} \label{alg_um}
  e^{-i \Delta s_i D_1} = 
\begin{pmatrix} 
1 & 0 & 0 \\
0 & e^{-i \Delta s_i \alpha} & 0  \\
0 & 0 & e^{-i \Delta s_i \gamma } 
\end{pmatrix}
\equiv U_m \ .
\end{equation}
Using Eqs. (\ref{alg_p} - \ref{alg_um}) one can iteratively find the S-matrix and the associated probabilities of Eq. (\ref{prob_s}). We will call this approach the iterations method. In the proof of Eq. (\ref{alg_s}) one needs the transformations forth and back between the flavor amplitudes and the mass eigenstates amplitudes,
\begin{equation} \label{flav_ch}
\nu_{\alpha} = \sum_i U_{\alpha i} \nu_i\ ;\ \nu_i\ = \sum_{\alpha} \left( U^{\dagger}\right)_{i \alpha} \nu_{\alpha} \ .
\end{equation}
The condition for small $\Delta s_i$ used in Eq. (\ref{alg_s}) suggests that one needs a large number of iteration to obtain good accuracy.  Our numerical implementation indicates that even 10-15 factors in Eq. (\ref{alg_p}) would provide an 0.1\% accuracy when compared with the ``exact" eigenvalues method. Fig. \ref{fig:errd} shows the difference between the iterations method described above when only 15 iterations are used, and the exact eigenvalues method solution. Increasing N to 150 reduces the absolute difference to less than $10^{-5}$.

\begin{figure}[htb]
\centering
\includegraphics[height=3.5in]{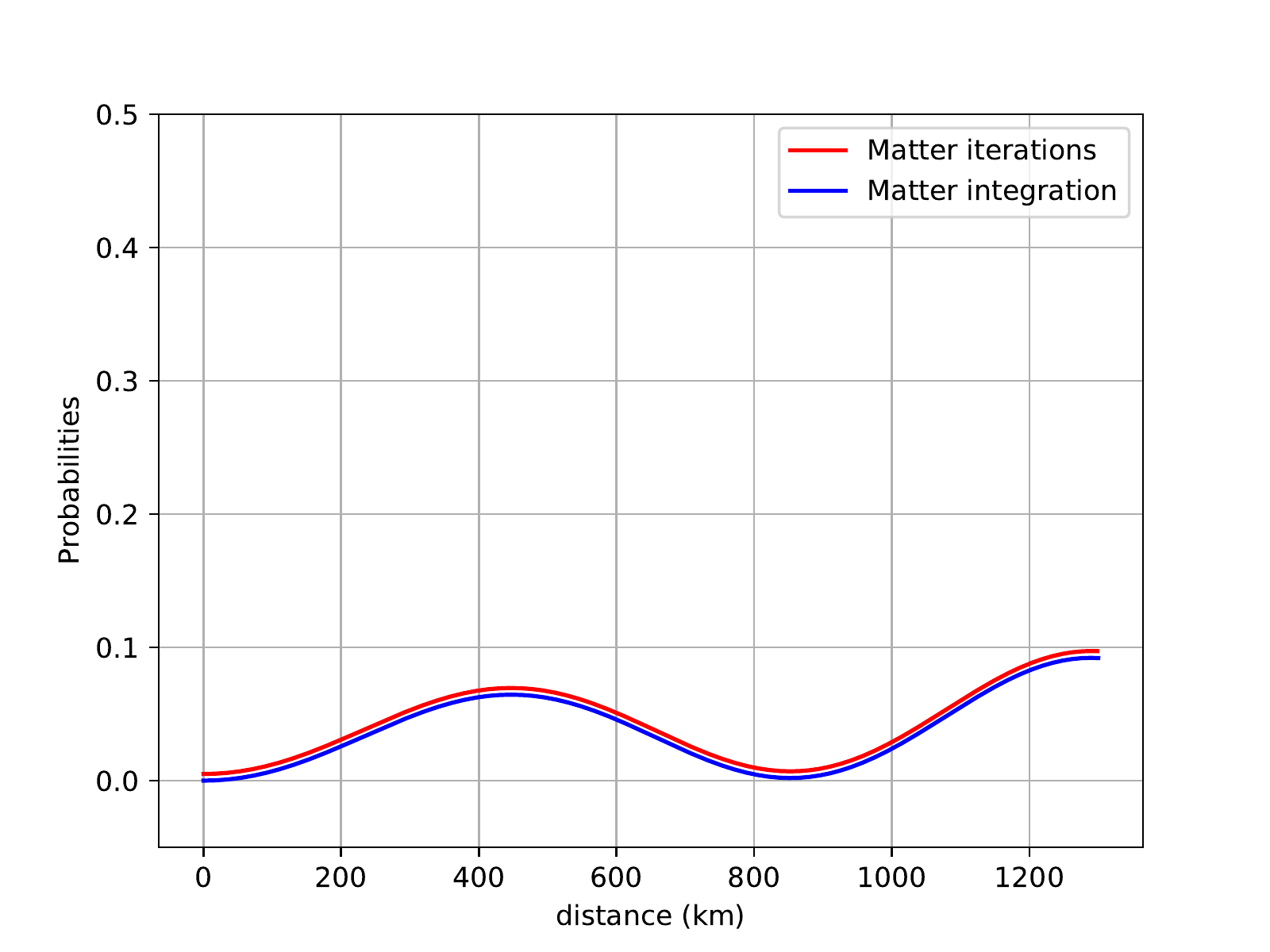}
\caption{Comparison of the electron neutrino appearance probability calculated with the iteration method described in the text and the direct integration method for smoothly varying electron (matter) density (see text for details). The two curves are artificially separated by 0.005 for a better view. A muon neutrino beam of 0.8 GeV was used in the calculations.}
\label{fig:iter}
\end{figure}

The iterations method described above works as well for smoothly varying average electron (matter) densities. Fig. \ref{fig:iter} shows the results of the above method (red curve) compared with the solution obtained by directly integrating Eq. (\ref{wolf}) for a varying density through the Earth crust similar to that described in Ref. \cite{ecrust17} (blue curve). Calculated is the electron neutrino appearance probability for a muon neutrino beam of 0.8 GeV. The two curves are overlapping if the artificial 0.005 shift to the red curved is removed.

The iterations algorithm described above can also be used to understand the results of section \ref{spikes}. To see that, one can consider the electron density in condensed matter composed of N spikes clustered around the atomic nuclei. One can further approximate the spikes with Dirac delta functions, which are normalized to unity, multiplied by the constant given in Eq. (\ref{norm_r}). Then,  when integrating Eq. (\ref{wolf}) on each $\Delta s_i$ segment, one can (i) transform the flavor amplitudes into mass-eigenstate amplitudes with $U^{\dagger}$ (see e.g. Eq. (\ref{flav_ch})), (ii) freely propagate the mass-eigenstate amplitudes using $U_m$, (iii) transform the mass-eigenstate amplitudes into flavor amplitudes with $U$, and (iv) integrate Eq. (\ref{wolf}) over the Dirac delta spikes. For the last step one can assume that in the vicinity of a Dirac delta spike only the term proportional with $A(s)$ in Eq. (\ref{wolf}) survives,
\begin{equation}
i \frac{d \nu_e (s)}{ds} = A(s) \nu_e (s) \ .
\end{equation}
Given that the Dirac delta function norm in Eq. (\ref{norm_r}) is proportionate to $\rho \left(s/N\right)=\rho \Delta s_i$, the solution to the above equation becomes
\begin{equation}
\nu_e(s_a) = e^{-i \Delta s_i A(s_i)} \nu_e(s_b)\ ,
\end{equation}
where $s_b$ and $s_a$ are the $s$ coordinates before and after the delta spike in the $\left(s_i,\ s_i+\Delta s_i \right)$ interval, and $A(s_i)$ is calculated with the average electron density of Eq. (\ref{rho_ave}).
Therefore, the result of integrating the full vector of amplitudes over the Dirac delta spike is
\begin{equation}
\begin{pmatrix} 
\nu_{e} \\
\nu_{\mu} \\
\nu_{\tau}
\end{pmatrix} (s_a)= 
\begin{pmatrix} 
e^{-i \Delta s_i A(s_i)} & 0 & 0 \\
0 & 1 & 0  \\
0 & 0 & 1   
\end{pmatrix}
\begin{pmatrix} 
\nu_{e} \\
\nu_{\mu} \\
\nu_{\tau}
\end{pmatrix} (s_b) \equiv U_m
\begin{pmatrix} 
\nu_{e} \\
\nu_{\mu} \\
\nu_{\tau}
\end{pmatrix} (s_b)\ .
\end{equation}
Putting together all the steps of the algorithm described above, one gets for the  S-matrix factors entering Eq. (\ref{alg_p})
\begin{equation} \label{alg_ss}
S(\Delta s_i) = U_A(s_i) U U_m U^{\dagger} \ ,
\end{equation}
which are the same as in Eq. (\ref{alg_s}). This completes the proof that justifies the use of an average electron density, rather than its large variation around the atomic nuclei. For smooth changes of the average electron density one can use a typical coarse-graining argument.

\section{Conclusion and Outlook}
In conclusion, we analyzed the effect of the large electron density variations around the atomic nuclei on the neutrino oscillation probabilities in condensed matter. The analysis is relevant for the DUNE/LBNF experiment. In section \ref{spikes} we attempted to fully integrate the evolution equation for the neutrino amplitudes, by considering the large variation of the electron density near the atomic nuclei, and therefore that of the neutrino potential. We found out that the numerical integration under these conditions could be treacherous, and could be leading to erroneous results.

In the second part of the manuscript we proposed a new iterative solution to the neutrino amplitudes evolution equation, which proves to be very fast, reliable, and applicable to either constant matter density or slowly varying matter density (assuming average electron densities). Finally, we showed that one can obtain the same iterative solution, by assuming that the spikes in the electron density around the atomic nuclei can be approximated by Dirac delta functions. Our solution thus justifies the use of average electron densities for matter effects in neutrino oscillation probabilities.

\section*{Acknowledgements}
AZ would like to thank the Central Michigan University Office of Research and Graduate Studies Summer Scholar Grant, Central Michigan University Department of  Physics. He also acknowledges collaboration with Marco Fornari of the AFLOW Consortium (http://www.aflow.org) under the sponsorship of DOD-ONR (Grants N000141310635 and N000141512266).  A special thank to Ethan Stearns for all his DFT calculations and help with Figure 1.



 
\end{document}